# Spatio-temporal evolution dynamics of ultrashort Laguerre-Gauss vortices in a dispersive and nonlinear medium

Shakti Singh[1], Akhilesh Kumar Mishra[2,*]

Department of Physics, Indian Institute of Technology Roorkee, Roorkee-247667, Uttarakhand, India
ssingh7@ph.iitr.ac.in[1], akhilesh.mishra@ph.iitr.ac.in[2]
*Corresponding author

**Abstract**

The additional degree of freedom as introduced by the orbital angular momentum (OAM) of light has revolutionized the various technological applications. Optical pulses with OAM have applications in the generation of twisted attosecond pulses, ultrafast spectroscopy, telecommunication, high harmonic generation and in many other areas of physics as well. In this paper, we present a numerical investigation of the propagation dynamics of ultrashort Laguerre Gauss (LG) vortices using nonlinear envelope equation (NEE) in a dispersive and nonlinear medium. Asymmetric splitting of ultrashort LG vortices in time at its bright caustic are observed above a certain pulse power. The asymmetric splitting owes its origin to space-time focussing. We also observe that spatial evolution at pulse center ($t = 0$ in pulse frame) and temporal evolution at bright caustic of the ultrashort LG vortex are quite similar. In the spectral domain, oscillatory structures are formed and above a certain peak power, we observe the generation of new frequency components with more intense lower frequency components.

## 1. Introduction

The fascinating properties of structured light has attracted a lot of research interest since last two decades. One aspect of structured beam is to carry orbital angular momentum (OAM). An optical beam which has azimuthally varying phase front ($e^{-il\Phi}$) in azimuthal direction $\Phi$, perpendicular to the propagation direction is known as vortex beam. In vortex beam phase is undefined at the centre of beam and optical field strength vanishes, which provides the beam with ring structure [1-5]. The integer $l$, called topological charge defines number of twist light does in one wavelength. Depending upon the handedness of phase front $l$ can take either positive or negative values and determines OAM magnitude as $l\hbar$ per photon of the beam [6-8]. Conventional vortex beams are Laguerre Gauss ($LG_l^p$) beams, which is the solution of Helmholtz equation in cylindrical coordinate system under paraxial approximation, where $l$ and $p$ respectively represent azimuthal index and number of radial modes in intensity distribution [9-10]. LG beams with different values of topological charge $l$ are orthogonal to each other, therefore it can provide additional degree of freedom to the photon in the beam. This additional degree of freedom has potential technological applications in optical communication [11], micromanipulation [12], phase-contrast microscopy [13], quantum information processing [14] and others [15].

[1]

Since its advent, the ultrashort pulsed laser has attracted a great deal of attention. These sources can govern the light in time duration up to femtosecond. As a result, ultrashort laser pulses have become essential for understanding the ultrafast phenomena [16]. In the ultrashort regime spatial and temporal properties are not independent of each other, rather there is a coupling between them called spatiotemporal coupling [17-19]. It means spatial and temporal properties of an ultrashort pulse are not independent of each other. Ultrashort pulses with OAM are known as ultrashort vortices. In ultrashort vortices there exist an additional coupling between azimuthal and temporal degrees of freedom due to which temporal shape strongly depends upon OAM, as described for ultrashort LG vortices and X waves [20-21]. Recently, it has been shown that for ultrashort LG vortices there exist an upper bound to the topological charge $l$ for a given pulse width and lower bound to pulse duration for a given $l$. Consequently, temporal pulse width with certain available spectral bandwidth cannot be independent of topological charge $l$ [22]. Setting a limit to the temporal width with OAM or a limit to the degree of freedom of OAM in ultrashort vortices has been employed in the diverse areas of physics such as in optical communication to limit the OAM based channel and hence it limits the communication capacity. These limits also have implications in ultraviolet high harmonic generation and quantum information, a limitation to multidimensional entanglement of OAM states [11-15]. These applications warrant investigation of spatio-temporal propagation dynamics of an ultrashort pulse with OAM in different kinds of dispersive and nonlinear media.

In this paper, we report modelling of the spatio-temporal evolution dynamics of ultrashort LG vortices in dispersive and nonlinear media. For modelling the propagation of ultrashort optical pulses, we solve the three-dimensional nonlinear Schrödinger (NLSE) equation. The employed NLSE assumes that the spectral content of the field is narrower than the carrier frequency and slowly varying envelope approximation (SVEA) is satisfied. However, when ultrashort optical pulses undergo self-focusing, self-phase modulation (SPM) causes to broaden the initial spectrum and therefore SVEA is no longer satisfied in the time domain. Therefore, to study the ultrashort optical pulses beyond SVEA, we include self-steeping and space-time focusing terms in the NLSE equation ,and after inclusion of these terms the new equation is named as nonlinear envelope equation (NEE) [23-24]. NEE describes accurately the propagation of ultrashort optical pulses up to single cycle regime. In the following section 2 details the theory and the model. Section 3 presents the results of the numerical model. This section discusses the temporal, spatial, and spectral evolution of the ultrashort LG vortex in detail. The last section, that is section 4, presents a summary of the work.

## 2. Theory and the model

Generally optical pulses are described as product of purely spatial and purely temporal terms. However, this condition fails in the case of broadband spectrum pulses and non-separable spatial and temporal condition comes into the play known as spatio temporal couplings. An optical pulse with spatio-temporal couplings upon propagation gives us special effects such as isodiffraction and isodivergence [17-19]. An optical pulses with OAM in ultrashort regime contain an addition couplings called azimuthal temporal couplings which makes the temporal shape of optical pulses strongly depends upon the OAM. These spurred a huge research interest in ultrashort pulses with OAM. For synthesizing the ultrashort optical vortices experimentally, we must overcome some technical limitations such as spatial, group velocity, and topological charge dispersions [25-27]. The ultrashort LG vortices of topological charge $l$ are synthesized by superposing LG beams of different frequencies and the same topological charge $l$. The analytical complex representation of LG vortices reads as

$$E(r,\phi,z,\tau) = \frac{1}{\pi}\int_0^\infty \hat{E}_\omega(r,\phi,z)e^{-i\omega\tau}d\omega \qquad (1)$$

where the superposed LG beams are given by

$$\hat{E}_\omega(r,\phi,z) = \hat{a}_\omega \frac{e^{-i(|l|+1)\Psi_\omega(z)}e^{-il\phi}}{\sqrt{1+\left(\frac{z}{z_{R,\omega}}\right)^2}} \left(\frac{\sqrt{2}r}{s_\omega(z)}\right)^{|l|} \frac{e^{i\omega\frac{\tau^2}{2cq_\omega(z)}}}{} \quad (2)$$

These beams carry same topological charge $l$ and zero radial order. In the equation (2) we have dropped the Laguerre Gauss polynomial $\left(L_p^{|l|}\left(\frac{2r^2}{s_\omega^2(z)}\right)\right)$ because for radial order $p=0$ the value of polynomial is always 1. $\hat{a}_\omega$ is the spectrum of LG beams. In equation (2) $\tau = t - \frac{z}{c}$ is the local time, $c$ is the speed of light in vacuum, $q_\omega(z) = z - iz_{R,\omega}$ is the complex beam parameter, $s_\omega(z) = s_\omega\sqrt{1+\left(\frac{z}{z_{R,\omega}}\right)^2}$ is the beam width, $s_\omega = \sqrt{\frac{2z_{R,\omega}c}{\omega}}$ is the waist of fundamental Gaussian beam at $z=0$ and $z_{R,\omega}$ is the Rayleigh range. The constant $\rho = \frac{r}{s_\omega(z)}$ represents revolution about axis $z$, called caustic surface [28]. It represents geometry of energy density distribution. Complex beam parameter $q_\omega(z)$ is generally written as

$$\frac{1}{q_\omega(z)} = \frac{1}{R_\omega(z)} + \frac{i2c}{\omega s_\omega^2(z)} \quad (3)$$

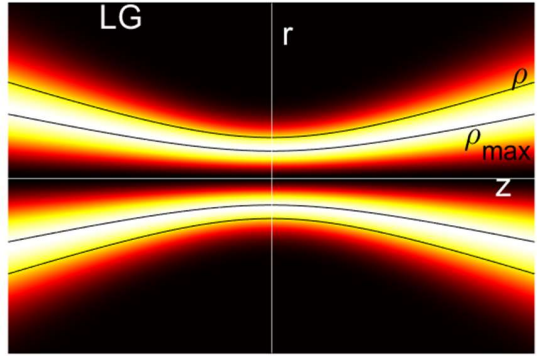

**Figure 1.** Cross section of a LG beam. Black curves represent revolution hyperboloids $\frac{r}{s_\omega(z)} = const$ about the $z$ axis, called here caustic surface of LG vortices. $\rho_{max}$ represents the bright caustic where the maximum energy density exists.

where $\frac{1}{R_\omega(z)} = \frac{z}{z^2+z_{R,\omega}^2}$ is the curvature of the phase fronts. For now, we assume that both the Rayleigh range $z_R$ and Gaussian waist width $s(z)$ are independent of the frequency [20].

The propagation of ultrashort LG vortices along $z$ direction in a dispersive medium with cubic Kerr type nonlinearity is modelled by NEE [23-24]

$$\frac{\partial A}{\partial z} = \frac{i}{2k}\left(1+\frac{i}{\omega}\frac{\partial}{\partial \tau}\right)^{-1}\nabla_\perp^2 A - \frac{i\beta_2}{2}\frac{\partial^2 A}{\partial \tau^2} + i\frac{n_2 n_o \omega}{2\pi}\left(1+\frac{i}{\omega}\frac{\partial}{\partial \tau}\right)|A|^2 A \quad (4)$$

where $k$ is propagation constant, $\beta_2$ is the group velocity dispersion (GVD) and $\tau = t - \frac{z}{c}$ is the local time. The presence of operator $\left(1+\frac{i}{\omega}\frac{\partial}{\partial \tau}\right)^{-1}$ in Laplacian term (first term on RHS) of equation (4) leads to space-time focusing while its presence with Kerr nonlinear term (third term on RHS) leads to self- steeping. Since the LG vortex has cylindrical symmetry, the

[3]

Laplacian in equation (4) is expressed as $\left(\frac{\partial^2}{\partial r^2} + \frac{1}{r}\frac{\partial}{\partial r} + \frac{1}{r^2}\frac{\partial^2}{\partial \phi^2}\right)$. For modelling equation (4) numerically we have made this equation dimensionless by adopting following transformations

$$u = \frac{A}{A_0}, \rho = \frac{r}{w_0}, T = \frac{\tau}{\tau_0}, Z = \frac{z}{L_{DF}} \qquad (5)$$

Using above transformations, we get following dimensionless form of NEE

$$\frac{\partial u}{\partial Z} = \frac{i}{4}\left(1 + \frac{i}{s}\frac{\partial}{\partial T}\right)^{-1} \nabla_\perp^2 u - i\frac{L_{DF}}{L_{DS}}\frac{\partial^2 u}{\partial T^2} + i\frac{L_{DF}}{L_{NL}}\left(1 + \frac{i}{s}\frac{\partial}{\partial T}\right)|u|^2 u \qquad (6)$$

where $L_{DF} = \frac{kw_0^2}{2}$ is diffraction length, $L_{DS} = \frac{t_0^2}{|\beta_2|}$ is dispersion length, $L_{NL} = \left(\frac{n_0 n_2 \omega |A_0|^2}{2\pi}\right)^{-1}$ is nonlinear length and $s = \omega\tau_0$ represents self-steeping parameter. The three characteristic lengths $L_{DF}$, $L_{DS}$ and $L_{NL}$ characterize the strengths of diffraction, dispersion, and nonlinear interactions of the LG vortices with the medium, respectively. The ratio of $\frac{L_{DF}}{L_{NL}}(= N^2)$ represents the strength of Kerr nonlinearity.

We consider the propagation of the ultrashort LG vortices in bulk fused silica. The functional form of the input LG vortices in space, time, and azimuthal direction is given as

$$A(r,t,\phi) = A_0 \exp\left(-\frac{\tau^2}{2\tau_0^2}\right) \exp\left(-\frac{r^2}{2s_0^2}\right)\left(\frac{r}{s_0}\right)^{|l|} \exp(il\phi) \qquad (7)$$

To see the evolution of LG vortices, we numerically integrate equation (6) employing the split-step Fourier transform method. We have considered the propagation in fused silica and used following simulation parameters: $\frac{L_{DF}}{L_{DS}} = 0.15$, $s = 110$, $\beta_2 = 385 fs^2/cm$ and $n_2 = 2 \times 10^{-6} cm^2/W$. We have considered $78fs$ LG vortex operating at $795nm$. We examine the spatial, temporal and spectral evolutions of the vortex pulse with propagation distance in the medium with aforementioned parameters. Moreover, we discuss the evolution at different strength of Kerr nonlinearity and for different values of the topological charge $l$. In our numerical experiment, we have considered the same energy ultrashort LG vortices of different topological charge $l$. To numerically achieve equal energy optical vortex pulses, first we calculated the energy $U$ of an ultrashort LG vortex by using $U = 2\pi \int_0^\infty r dr \int_{-\infty}^\infty |A|^2 dt$. Assuming energy for a vortex pulse with $l = 2$ is given by $U_2$ and that for $l = 4$ is given by $U_4$, to keep the energy same in both the cases, we scaled $U_4$ such that the energy for ultrashort vortex with $l = 4$ becomes $U_4/\left(\frac{U_4}{U_2}\right)$, which is same as $U_2$. Such a scaling mimics experimental scenario where we launch a pulse of fixed energy on spatial light modulation, which in turn generates vortex pulses with varying OAM but of same assumingly energy.

The paper is organised as follows- in section (3.1) we discuss the temporal evolution of LG vortices at the bright caustic of LG vortices. Section (3.2) discusses the spatial evolution of LG vortex at time $(t = 0)$ in pulse frame. Section (3.3) investigates the spectral evolution at the bright caustic of LG vortices and in the last section (3.4) we study the evolution of maximum peak intensity of ultrashort LG vortex with propagation.

### 3. Results and discussion

3.1 *Temporal Evolution*

In this section, we investigate the temporal evolution of a pulsed LG vortex at its bright caustic. Fig. 2(a) shows the pulse evolution for $l = 2$. We observe that with propagation initially the peak intensity of the pulse increases and thereafter the pulse broadens. The temporal focusing appears due to the stronger SPM effects as $\frac{L_{DF}}{L_{DS}} = 0.15$ and $\frac{L_{DF}}{L_{NL}} = 5.8$. At larger distances, dispersion takes over and therefore, because of broadening in time, peak intensity starts to decrease. The pulse evolution for $l = 4$ is shown in fig. 2 (b). Here we see that the pulse compression is relatively weak, which results in smaller peak height as compared to that in fig. 2 (a). This is because of the larger value of the topological charge $l$ as larger $l$ reduces the input peak intensity of the LG vortex, which in turn leads to weaker nonlinear effects. For larger input power pulse, that is, for larger value of $N^2$, the pulse temporal evolution for topological charges $l = 2$ and $l = 4$ are shown in figs. 2(c) and 2(d) respectively. Due to larger input power of the pulsed vortex beam, we observe temporal splitting after focussing in fig. 2 (c). The observed splitting evolves asymmetrically (trailing pulse have higher amplitude than leading pulse) due to space-time focusing [29]. For $l = 4$, the dynamics is shown in fig. 2 (d). Here we do not see pulse splitting at the same power because at higher topological charge peak intensity get reduced. If we further increase the strength of nonlinearity, we observe an early splitting and reversed evolution of the temporal splitting.

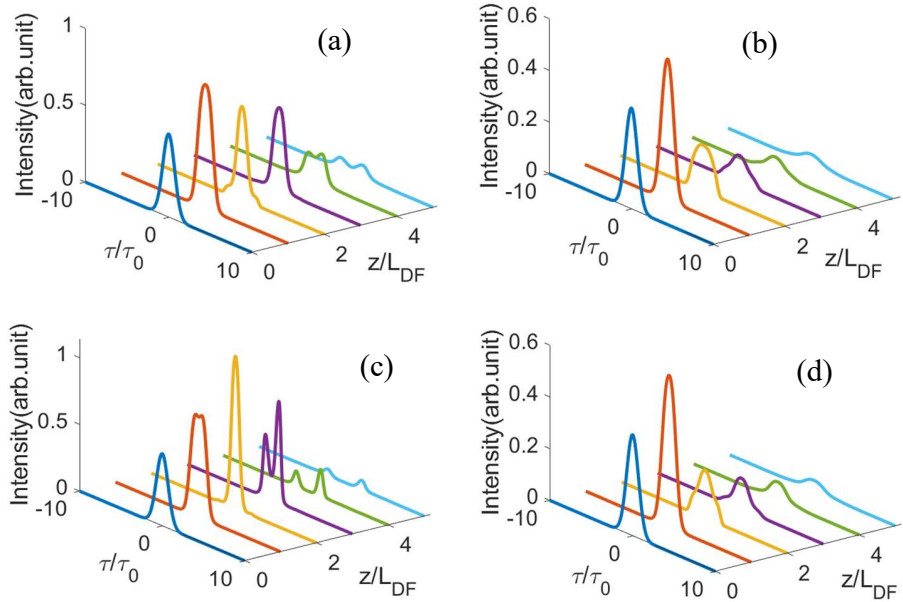

**Figure 2.** Temporal evolution of ultrashort LG vortices at the bright caustic. First row describes the evolution for $N^2 = 5.8$ for topological charge (a) $l = 2$ and (b) $l = 4$. Second row describes the evolution at $N^2 = 6.8$ for topological charge (c) $l = 2$ and (d) $l = 4$.

### 3.2 *Spatial Evolution*

In this section, we examine the spatial evolution of pulsed LG vortices at ($t = 0$) in pulse frame. The spatial evolution for $l = 2$ is depicted in fig. 3 (a). The figure shows that the pulsed LG vortex gets focussed with propagation. The focussing happens due to the combined effect of Kerr nonlinearity and diffraction and on further propagation the peak intensity decreases because diffraction comes into play, and this reduces the peak intensity of pulsed LG vortex. At higher value of topological charge peak intensity would be less and therefore pulsed LG vortex starts to broaden even from a relatively smaller propagation distance as shown in fig. 3(b), where $l = 4$. For larger value of the input power ($N^2 = 6.8$) Kerr effect dominates and hence the LG vortex remains focussed up to larger distance as shown in fig. 3(c). For $l = 4$ and $N^2 = 6.8$, fig. 3(d) depicts the spatial evolution. As clear from the figure, the pulsed LG vortex experiences faster spatial broadening. We also observe that for stronger nonlinearity, pulsed LG vortex gets focussed at smaller distance. Splitting at wings is observed in spatial evolution too as is visible in fig. 3(c) and 3(d) around $\frac{z}{L_{DF}} = 1$.

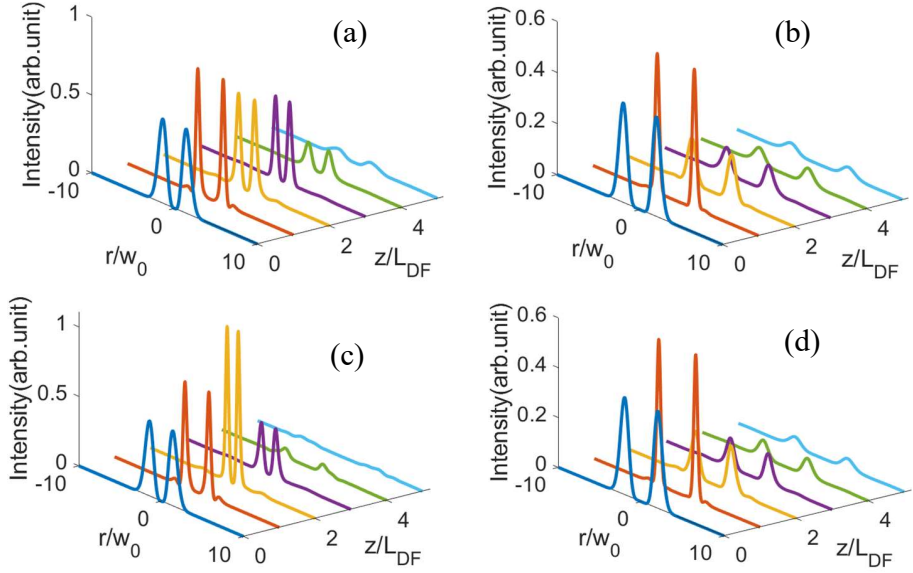

**Figure 3.** Spatial evolution of ultrashort LG vortices at time ($t = 0$) in pulse frame. First row describes the evolution at $N^2 = 5.8$ for topological charge (a) $l = 2$ and (b) $l = 4$. Second row describes the evolution at $N^2 = 6.8$ for topological charge (c) $l = 2$ and (d) $l = 4$.

Shown in Fig. (4) are spatial image profiles (surface plots) of spatial evolution of ultrashort LG vortices for on axis temporal profile for $N^2 = 6.8$. The first column represents the evolution of topological charge $l = 2$ and second column represent the evolution of topological charge $l = 4$. Figs. (a) and (b) represent the input of pulsed LG vortices in space, (c) and (d) represent the spatial profiles at $\frac{z}{L_{DF}} = 2$, and (e) & (f) at $\frac{z}{L_{DF}} = 4$. These figures clearly depict the ring evolution of formation of multi ring structure (ring splitting) as explained earlier.

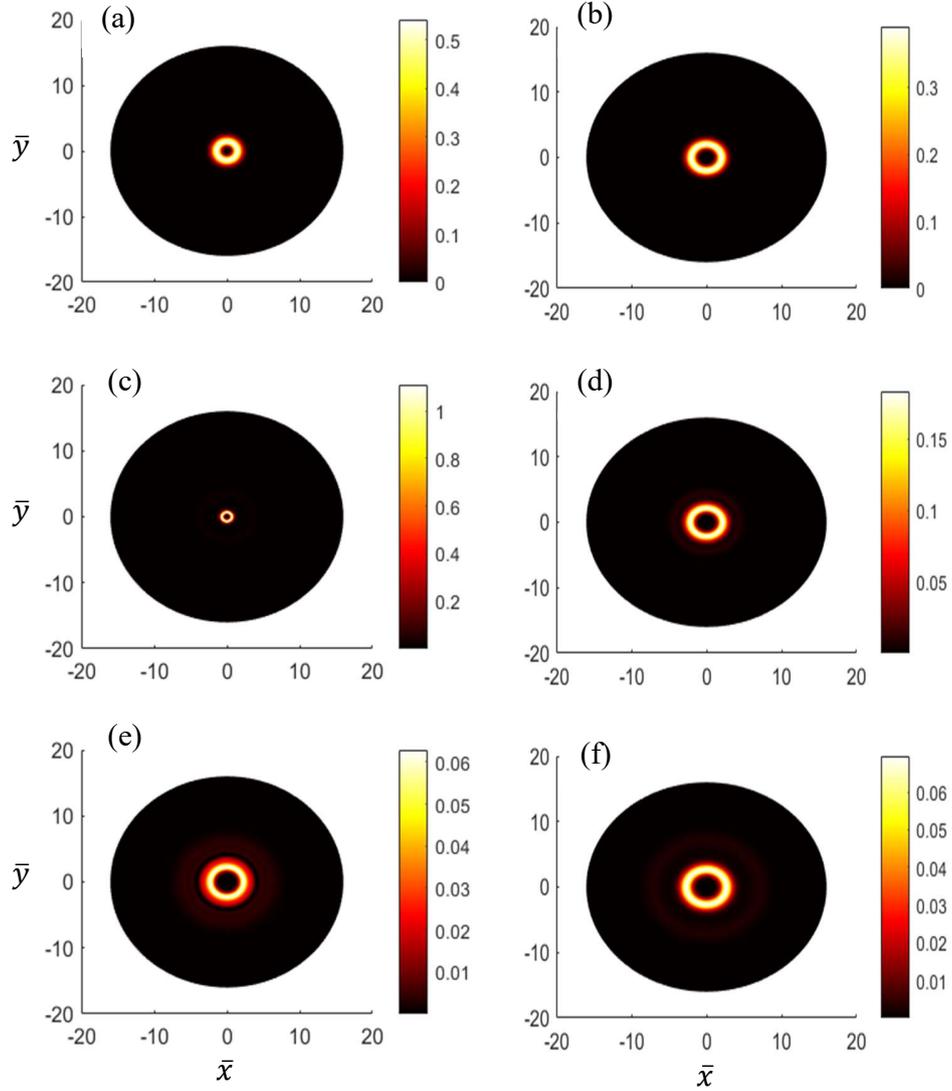

**Figure 4.** Surface plots of spatial evolution of ultrashort LG vortex at time ($t = 0$) in pulse frame. $\bar{x}$ and $\bar{y}$ represent the dimensionless coordinates. Evolution has been observed for $N^2 = 6.8$. The first column represents the evolution of topological charge $l = 2$ and second column represent the evolution of topological charge $l = 4$. (a) and (b) represent the input of LG vortices in space, (c) and (d) represent the spatial profile at $\frac{z}{L_{DF}} = 2$, (e) and (f) at $\frac{z}{L_{DF}} = 4$.

3.3 *Spectral Evolution*

[7]

Now we study the spectral evolution with the propagation of pulsed LG vortex at the bright caustic. As depicted in fig. 5 (a), the pulsed LG vortex spectrum for $l = 2$ broadens due to SPM induced phase shift $\Phi_{NL}$ [30]. Spectral asymmetry at $\frac{z}{L_{DF}} = 1$ is observed due to self-steeping in time where we observe that red-shifted peak is more intense compared to blue-shifted part. At larger distance the pulse depletes in time that reduces self-steeping and symmetric evolution is observed subsequently. Similar spectral evolution is also observed for $l = 4$ pulsed LG vortex as shown in fig. 5(b). Since for larger $l$, the peak intensity is relatively low, nonlinear effects are relatively weaker. For higher input power pulses of different topological charges are shown in figs. 5(c) and 5(d). In figure 5(c), due to stronger nonlinear effects, we see a distinct and stronger asymmetry at $\frac{z}{L_{DF}} = 1$, which is followed by self-phase modulation induced oscillations. Fig 5(d) shows the spectral evolution for $l = 4$.

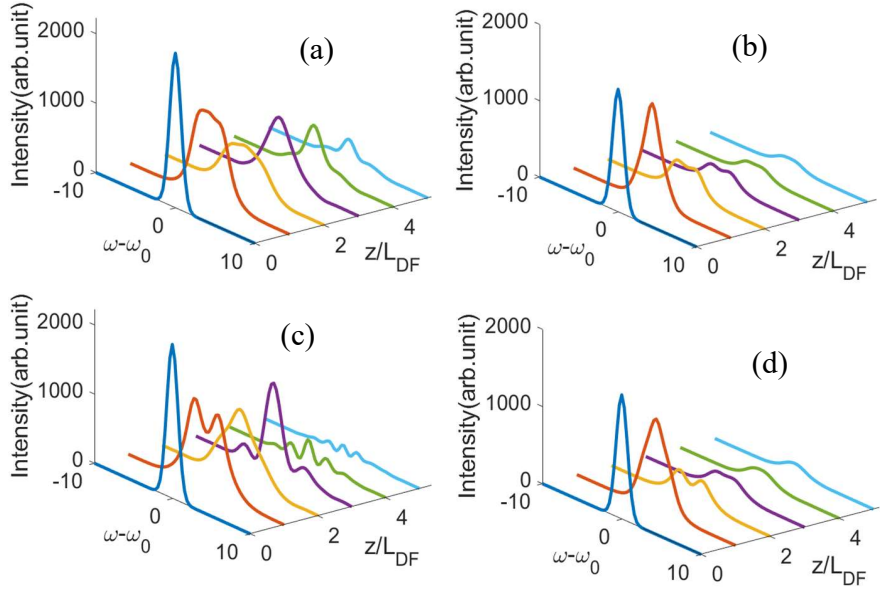

**Figure 5.** Spectral evolution of ultrashort LG vortices at bright caustic. First row describes the evolution at $N^2 = 5.8$ for topological charge (a) $l = 2$ and (b) $l = 4$. Second row describes the evolution at $N^2 = 6.8$ for topological charge (c) $l = 2$ and (d) $l = 4$.

### 3.4 *Temporal peak intensity evolution dynamics*

The evolution of temporal peak intensity of pulsed LG vortices is quite interesting as depicted in fig. 6.

We observe from fig. 6 that initially the temporal peak intensity increases with propagation due to Kerr nonlinearity induced focusing but after certain distance of propagation dispersion take over and that decreases the temporal peak intensity. For high power input pulses (see fig. 6(c) and fig. 6(d)), not only focusing appears earlier but pulse broadens relatively faster. Shift in the peak intensity for larger $l$ is also evident in the figure.

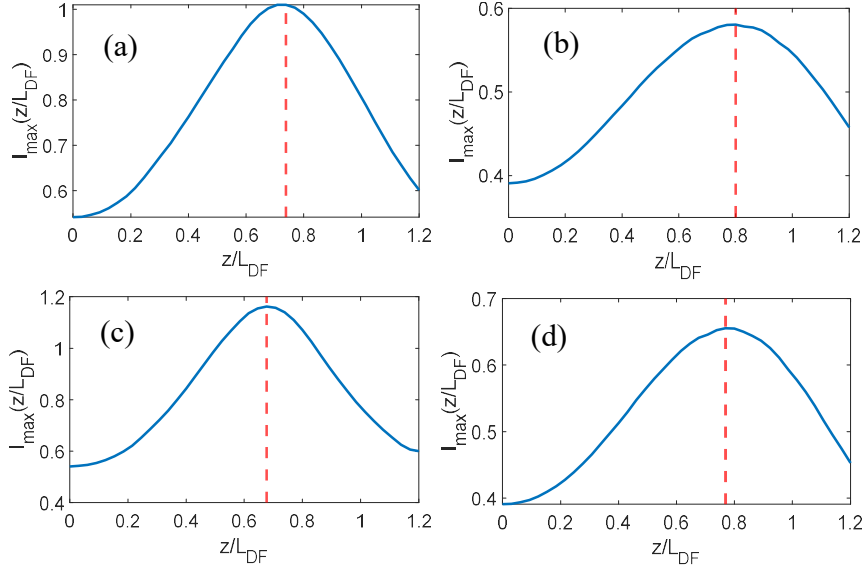

**Figure 6.** Evolution of temporal peak intensity of the ultrashort LG vortex at the bright caustic. First row describes the evolution at $N^2 = 5.8$ for topological charge (a) $l = 2$ and (b) $l = 4$. Second row describes the evolution at $N^2 = 6.8$ for topological charge (c) $l = 2$ and (d) $l = 4$.

### 4. Conclusion

In summary we have numerically studied the spatio-temporal evolution dynamics of ultrashort LG vortices in bulk silica medium with the help of NEE. The numerical experiments mimic the experiments performed with same energy pulses with different topological charge $l$. For same energy pulses, larger $l$ results reduced peak intensity. In temporal domain, pulse with smaller $l$, due to stronger nonlinearity, undergoes stronger compression. In addition, owing to the space time focussing asymmetrical splitting has also been observed. For pulse with larger $l$ temporal evolution observes weaker compression. Similar effects have also been observed during spatial evolution (at $t = 0$) in pulse frame. Further, in spectral domain, we have observed Kerr-nonlinearity and self-steeping effect induced spectral features. Furthermore, we have studied the evolution of temporal peak intensity of the ultrashort LG vortex at the bright caustic.

The study may find applications in optical communication where we can limit the OAM based multiplexing channel, in compression of femtosecond LG vortices, supercontinuum generation of ultrashort LG vortices, temporal resolution of ultrafast spectroscopy, optical tweezers of tunable OAM etc.

### Acknowledgement

Shakti Singh wishes to acknowledge the University Grant Commission (UGC) India for the financial support.

[9]